\documentclass[aps,prb,twocolumn,showpacs]{revtex4}
\usepackage{graphicx}
\usepackage{psfrag}
\usepackage{color}
\usepackage{amsmath}
\usepackage{bm}

\begin{document}

\title{Predicted light scattering by Josephson vortices in layered cuprates}

\author{O.~V.~Dimitrova$^{+}$\/\thanks{e-mail: olgdim@itp.ac.ru} and T.~M.~Mishonov$^{*}$}
\affiliation{$^+$CNRS, Universit\'e Paris-Sud, UMR 8626, LPTMS, Orsay Cedex, F-91405 France
\\~\\
$^*$Department of Theoretical Physics, Faculty of Physics, University of Sofia St. Clement of Ohrid,
5 J. Bourchier Boulevard, Bg-1164 Sofia, Bulgaria}


\begin{abstract}
Light scattering by Josephson vortices is theoretically predicted. An experimental set-up
for the observation of the effect is proposed. The magnetic field should be parallel to the
CuO$_2$ layers and the crystal surface. The falling light should be at grazing angles,
and the Bragg scattering must be almost backscattering. The distance between
the columns of the Josephson vortices must be slightly smaller than the light-wavelength.
The surface should be very clean and flat in order light scattering by surface terraces
and surface roughness to give the relatively small background.
\end{abstract}

\pacs{71.23.An, 72.15.Rn, 72.25.Ba, 73.63.Nm }

\maketitle

\section{Introduction}
The purpose of the present work is to attract the attention of the experimentalists
to the possible existence of a new effect in the physics of the layered high-$T_c$ cuprates --
the light scattering by the Josephson lattice as by a grating made from a non-transparent material.

\section{Model}

For the description of a typical layered superconductor like the high-temperature superconductor (HTSC)
$\mathrm{YBa}_2 \mathrm{Cu}_3\mathrm{O}_7$
we use the Lawrence-Doniach (LD) model~\cite{LD} of the Josephson-coupled superconducting layers.
In this model the material is treated as a stack of superconducting planes, separated by
an insulating material. When the spacing between the layers tends to zero,
the effects of the microscopic layered structure are averaged out, and the LD model
reduces to the anisotropic Ginzburg-Landau model. In the latter the anisotropic
nature of the material appears only in the form of a mass tensor with unequal principal values.

The LD Gibbs free energy of the superconducting layers parallel to the $xy$ plane and stacked
along the $z$-axis is given by
\begin{eqnarray}\label{FLD}
{\cal{F}}_{_{\mathrm{LD}}}&=&c_0 \int d^2\vec{\rho} \sum_{n=0,\pm 1,..} \!F_n \\
&&+\int\frac{1}{8\pi}\left[\mathrm{rot} \,{\mathbf A}({\mathbf r})-{\mathbf  B}({\mathbf r})\right]^2dx\,dy\,dz,
\nonumber
\end{eqnarray}
where $c_0$ is the spacing between the superconducting layers,
$\vec{\rho}=(x,y)$
is the two-dimensional in-plane radius-vector.
The superconducting layers are enumerated by an integer $n$, with the coordinate of the $n$-th layer
$z_n=nc_0$ (see, for instance, Fig.~4 in Part II of Ref.~[\onlinecite{Kopnin}])
\begin{eqnarray}
{\mathbf A}({\mathbf r})=\left(\begin{array}{r}
\vec{A}\,({\mathbf r})   \\
A_z({\mathbf r})\end{array}\right)\qquad\mbox{and}\qquad{\mathbf r}=\left(\begin{array}{l}
\vec{\rho}   \\
z\end{array}\right)
\end{eqnarray}
are the three-dimensional vector-potential and space-vector. ${\mathbf  B}({\mathbf r})$ is the
external magnetic field.

In the representation when the order parameter is the wave function of the superconducting electrons,
the free energy density $F_n$ of the $n$-th superconducting layer from the LD functional Eq.~(\ref{FLD}) is
given by
\begin{eqnarray}\label{Fn}
&&F_n=\alpha(T)\left|\Psi_n(\vec{\rho})\right|^2+\frac{\beta(T)}{2}\left|\Psi_n(\vec{\rho})\right|^4 \\
&&+\frac{1}{2m^*}\left|\left(-i\hbar\nabla_{_{\mathrm{\!\!2D}}}
-\frac{e^*}{c}\vec{A}_n(\vec{\rho})\right)\Psi_n(\vec{\rho})\right|^2
\nonumber \\
&&+\frac{\hbar^2}{M c_0^2}\!
\left[1\!-\!\cos{(\theta_{n+1}-\!\theta_n-\!\frac{e^*}{\hbar c}\!\int_{nc_0}^{(n+1)c_0}\!\!\!\!A_zdz)}\right]
\!\left|\Psi_n(\vec{\rho})\right|^2,
\nonumber
\end{eqnarray}
where $\alpha(T)$ and $\beta(T)$ are
Ginzburg-Landau parameters, which are constant in space but depend on temperature, $\Psi_n(\vec{\rho})$  and
$\theta_n(\vec{\rho})$ are the wave function and the superconducting phase only defined on the $n$-th plane,
$M$ and $m^*$ are the effective masses of the Cooper pairs
for the motion along the z-direction and in the $xy$-plane (the coordinate axes were chosen in such a way,
that they coincide with the
principal directions of the mass tensor of the Cooper pairs),  $e^*=2e$ is the charge of the Cooper pairs,
$\nabla_{_{\mathrm{\!\!2D}}}$ is the 2D gradient in the
$xy$ plane, $\vec{A}_n(\vec{\rho})=\vec{A}(\vec{\rho},nc_0)$, $n=0,\pm 1,..$,
is the restriction of $\vec{A}$ to the $n$-th superconducting plane.
The presence of the vector-potential in the free energy, Eq.~(\ref{Fn}), is due
to the gauge invariance, the phase of the order parameter has to come in the combination
$\theta-e^*/\hbar c\int {\mathbf A}\cdot d{\mathbf r}$.
Thus enclosed in the round brackets in the third line of Eq.~(\ref{Fn})
is the gauge-invariant phase difference.
This is also the discrete analogue of the corresponding gradient term.

The material is isotropic and homogeneous
in planes parallel to the superconducting layers.
However, in the perpendicular direction the superconducting behavior is governed by the anisotropy parameter
\begin{equation}\label{Gamma}
\Gamma=\sqrt{\frac{M}{m^*}}.
\end{equation}

The third line in Eq.~(\ref{Fn}) describes the Josephson interaction between the layers.
We consider the case of a extremely anisotropic superconductor $\xi_c\ll c_0$, where
$\xi_c(T)=\hbar/\sqrt{2m^*|\alpha(T)|}$
is the coherence length transverse
to the layers direction (when it exists in the continuous limit of $c_0\rightarrow 0$). In this case, since
the center of a vortex parallel to the layers is located in the middle of the two adjacent  planes, both
SCsuperconducting planes are far from the normal vortex core of size $\xi_c$ in the transverse direction,
which is supposed to exist in the continuous limit.
Therefore, in the third line of the Josephson interlayer coupling term Eq.~(\ref{Fn}),  for the Cooper pair wave function
was substituted  its equilibrium value $\Psi_n(\vec{\rho})=\Psi_{_{\mathrm{\!GL}}} e^{i\theta_n (\vec{\rho})}$, where
the modulus is
\begin{equation}\label{Psigl}
\Psi_{_{\mathrm{\!GL}}}=\sqrt{\frac{|\alpha|}{\beta}},
\end{equation}
and roughly equals the average superconducting density.

Variation of the LD functional, Eq.~(\ref{FLD}), with respect to the parallel and the transverse components of the vector-potential
$\delta {\cal F}/\delta \vec{A}_n=0$ and $\delta {\cal F}/\delta A_z=0$ , gives the in-plane current
\begin{equation}\label{jn}
j_n=-\frac{c}{4\pi\lambda_{ab}^2}\left(A_n-\frac{\hbar c}{e^*}\nabla_{_{\mathrm{\!\!2D}}}\theta_n\right),
\end{equation}
with the total current in the perpendicular to the $z$-direction given by
$j_{xy}=c_0\sum_n\delta(z-nc_0)j_n$, and the Josephson current between the $n$-th and the $(n+1)$-st layers
\begin{equation}\label{jz}
j_z=j_J\sin{(\theta_{n+1}-\theta_n-\frac{e^*}{\hbar c}\int_{nc_0}^{(n+1)c_0}\!\!\!A_zdz)},
\end{equation}
where $j_J=(c/4\pi\lambda_{c}^2) (\hbar c/e^*)/c_0$
is the Josephson critical current, and
$\lambda_{ab}=\sqrt{m^* c^2/4\pi n {e^*}^2}$ and $\lambda_c=\lambda_{ab}\Gamma$ are the
London penetration depths ($n\approx \Psi_{_{GL}}$).

Variation of the LD functional, Eq.~(\ref{FLD}), with respect to the phase
$\delta {\cal F}/\delta \theta_n=0$, gives the charge conservation law
$\nabla_{_{\mathrm{\!\!2D}}}
j_n+\Big(j_z^{n,n+1}-j_z^{n-1,n}\Big)/c_0=0$. Using the latter and
Eqs.~(\ref{jn}, \ref{jz}), one can obtain the vortex-conditions for the
gauge invariant phase difference. A typical  Josephson vortex in a layered compound  is
shown in Fig.~19 in Part II of Ref.~[\onlinecite{Kopnin}].
The magnetic field is along the $y$-axis.
The vortex is flattened out along the anisotropy $z$-axis.
The structure of the vortex depends on the relation between the London penetration
depths $\lambda_{ab},\lambda_c$ and the so-called Josephson screening length
$\lambda_J=c_0 \Gamma$.

\subsection{Continuous approximation}
In the continuous limit
of $c_0\rightarrow 0$ the expression in the square brackets in the third line of Eq.~(\ref{Fn}) transforms to
$$\frac{c_0^2}{2\hbar^2}\left|\left(-i\hbar\frac{\partial}{\partial z}-\frac{e^*}{c}A_z\right)\Psi_n(\vec{\rho})\right|^2,$$
and correspondingly the Lawrence-Doniach functional, Eq.~(\ref{FLD}), reduces to the anisotropic Ginzburg-Landau,
or effective mass, functional.  Variation of the anisotropic Ginzburg-Landau functional
(at a zero external field and a zero vector-potential) gives the static Ginzburg-Landau equation for uniaxial superconductor
\begin{eqnarray}\label{SchroedingerAnisotropic}
&&\frac{\delta {\cal F}}{\delta \Psi^*_n}=\\
&&-\frac{\hbar^2}{2 m^*}\left(\Delta_{_{\mathrm{2D}}}
+\frac{m^*}{M}\partial_z^2\right)\Psi_n
+\alpha\Psi_n+\beta|\Psi_n|^2 \Psi_n=0,\nonumber
\end{eqnarray}
where $\Delta_{_{\mathrm{2D}}}=\partial_x^2+\partial_y^2$, and
\begin{equation}
\frac{1}{c_0^2}\left[\Psi_{n+1}(\vec{\rho})-2\Psi_n(\vec{\rho})+\Psi_{n-1}(\vec{\rho})\right]\nonumber
\xrightarrow{c_0\rightarrow 0} 
\partial_z^2\Psi_{_{\!n}}
\end{equation}
is the continuous limit of the discrete second derivative.

\subsection{Isotropization for extremely type-II superconductors}
After the rescaling
\begin{equation}\label{rescaling}
\tilde{z}= z\Gamma,
\end{equation}
with the anisotropy parameter  $\Gamma$ from Eq.~(\ref{Gamma}),
the anisotropic Eq.~(\ref{SchroedingerAnisotropic}) transforms to
the isotropic Ginzburg-Landau equation
\begin{equation}\label{SchroedingerIsotropic}
\xi_c^2\Delta\Psi+\Psi-\frac{|\Psi|^2}{\Psi_{_{\mathrm{\!GL}}}^2}\Psi=0,
\end{equation}
with the Laplacian
\begin{equation}\nonumber
\Delta_{_{\mathrm{2D}}}+\frac{m^*}{M}\partial_z^2
\quad
\xrightarrow {\tilde{z}=\Gamma z}
\quad \Delta=\partial_x^2+\partial_y^2+\partial_{\tilde{z}}^2
,
\end{equation}
and $\Psi_{_{\mathrm{\!GL}}}$ being the modulus of the superconducting order parameter for
vortexfree space homogeneous phase by Eq.~(\ref{Psigl}) and $\xi_c$ is the coherence length in the z-direction.

The solution of Eq.~(\ref{SchroedingerIsotropic}) in the case of a vortex filament
with an axis parallel to the $y$-axis
is well known
\begin{equation}\label{ContinuousSolution}
\Psi=\Psi_{_{\mathrm{\!GL}}} \,\,f\left(\frac{\sqrt{x^2+\tilde{z}^2}}{\xi_c}\right) e^{i\phi},
\end{equation}
where $\sqrt{x^2+\tilde{z}^2}$ is the distance from the axis of the filament,
$\phi=\arctan (\tilde{z}/{x})$ is the
polar angle round the axis, and $f(\eta)$ is found from the equation
\begin{equation}\label{f}
\frac{1}{\eta}\frac{\partial}{\partial \eta}\left(\eta\frac{\partial f}{\partial \eta}\right)-\frac{f}{\eta^2}+f-f^3=0.
\end{equation}
The plot of the function $f(\eta)$  is shown in Fig.~4 in paragraph 30 of Ref.~[\onlinecite{LandauIX}]
$f(0)=0$ and $f(\infty)=1$. Thus, in the continuous limit and in the curved space $(x,\tilde{z})$, the vortex is round and
the coherence length $\xi_c$ is the radius of its normal core.

However, in our space $(x,z)$ the vortex is oblate.
Also, one has to take into account the discrete nature of the material.
For a qualitative evaluation
of the effect considered in this Letter, we use a trial function which is exact for the isotropic extremely type-II superconductors.
To obtain this trial function, one substitutes the rescaled discrete coordinates $\tilde{z}= (n-1/2)c_0\Gamma$
of the planes in the solution~(\ref{ContinuousSolution}) of the Ginzburg-Landau equation for isotropic superconductor
\begin{equation}\label{Trial}
\Psi_n(x)=R_n(|x|)e^{i\theta_n(x)},
\end{equation}
where $R_n(|x|)=\Psi_{_{\mathrm{\!GL}}}\,\,f(\sqrt{x^2
 +(n-1/2)^2c_0^2\Gamma^2}/\xi_c)$ with $f(\eta)$ defined by Eq.~(\ref{f}), and
\begin{equation}\label{thetan}
\theta_n(x)=\arctan\left[\frac{(n-1/2)c_0\Gamma}{x}\right]
\end{equation}
being the non-homogeneous superconducting phase in the $n$-th plane.

\section{Elastic light scattering}
Now let us analyze the Josephson current induced by the high-frequency vector-potential of the scattered light.
The static vector-potential of the Josephson vortex has a negligible influence, and the Josephson formula for the
current Eq.~(\ref{jz}) gives
\begin{equation}\label{ExcitedByLight}
j_z^{(\mathrm{in})}\approx -\frac{c}{4\pi\lambda_{c}^2}\cos{\left(\theta_{n+1}-\theta_n\right)}A_z^{(\mathrm{in})},
\end{equation}
where $A_z^{(\mathrm{in})}$ is the $z$-component of the vector-potential of the light
\begin{equation}\label{A}
{\mathbf A}^{(\mathrm{in})}({\mathbf r},t)={\mathbf A}_{\omega,k}^{(\mathrm{in})}e^{i{\mathbf k}{\mathbf r}-i\omega t},
\end{equation}
polarized in a plane perpendicular to the axis of the vortex filament
\begin{equation}\label{k}
{\mathbf k}=\left(k_x^{\mathrm{(in)}},0,k_z^{\mathrm{(in)}}\right).
\end{equation}

For a vortex-free phase $\theta_n=const$ we arrive at the London current response
$j_z^{(\mathrm{in})}=- (c/4\pi\lambda_{c}^2) A_z^{(\mathrm{in})}$.
Subtracting this homogeneous response, we derive the
net contribution from the Josephson vortex
\begin{equation}\label{ExcitedByLightNetVortex}
\delta j_z^{(\mathrm{in})}\approx \frac{c}{4\pi\lambda_{c}^2}
\left[1-\cos{\left(\theta_{n+1}-\theta_n\right)}\right]A_z^{\mathrm{(in)}}.
\end{equation}
The substitution here of the trial function for the phase distribution, Eq.~(\ref{thetan}), after some algebra gives
\begin{equation}\label{ExcitedByLightNetVortexTheta}
\delta j_z^{(\mathrm{in})}(x)\approx \frac{c}{4\pi\lambda_{c}^2}\,\,
\frac{2 (c_0 \Gamma/2)^2}{(c_0 \Gamma/2)^2+x^2}\,\,A_z^{(\mathrm{in})}
\end{equation}
for the current between layers $n=0$ and $n=1$.
The vortex ``core'' is ``sandwiched'' between those layers.
Supposing that the vector-potential
is approximately homogeneous at distances of the order of $c_0\Gamma$,
the integration of Eq.~(\ref{ExcitedByLightNetVortexTheta})
with respect to $x$ (using
$\int_{\infty}^{\infty}\frac{dx}{a^2+x^2}=\frac{\pi}{a}$) gives the $\delta$-like approximation
\begin{equation}\label{ExcitedByLightDelta}
\delta j_z^{(\mathrm{in})}(\vec{r})\approx \frac{c}{4\pi\lambda_{c}^2}\,\,
\Delta S \,\,\delta(x)\delta(z)\,\,
A_z^{(\mathrm{in})},\nonumber
\end{equation}
where $\Delta S=\pi c_0^2 \Gamma$. This formula has a very simple interpretation a single vortex operates as a
small tunnel parallel to the $y$-direction of the external magnetic field, with a cross-section $\Delta S$.

Now, summing the influence of all the vortices from the Josephson lattice
\begin{equation}\label{jzdeltadelta}
j_z^{\mathrm{(in)}}(x)=\frac{c}{4\pi \lambda_c^2}\,\Delta S \sum_{m=0,\,\pm 1, \,\dots}\delta(x-a m)\,\delta(z)\,\,A_z^{\mathrm{(in)}}(x).
\end{equation}
This lattice is like an extended
Abrikosov lattice along the $x$-direction. For a strong anisotropy $\lambda_c/\lambda_{ab}=\Gamma\gg 1$
the extended lattice of triangles looks like vortex columns spaced by distance
\begin{eqnarray}\label{aB}
a=2\Gamma\sqrt{\frac{\Phi_0}{\sqrt{3}B}},
\end{eqnarray}
where $\Phi_0=2\pi\hbar c/e^*$ is the flux quantum. In vertical direction the distance between two adjacent
vortices in one column is $b=\sqrt{3}a/\Gamma^2$. This distance is much smaller than the wavelength,
and if we wish to analyse the light scattering, we have to average the current response over the vortices
in one column. A single vortex operates as a hole with an area $\Delta S$, and a vortex column is equivalent
to a slit with width $w=\Delta S/b$. In such a way our electrodynamic problem is to calculate the intensity
of  light diffracted by regular grating slits with width $w$. In its depths the material is not transparent, that is why
the light can see only the top end of the vortex columns with a height $H<\lambda_{\mathrm{light}}$.
For an order of magnitude evaluation of the effect one can take $H\sim 100$~nm. Along this effective depth of the slits
one can have many vortices. At high enough magnetic field the distance between the Josephson vortices
can be of the order of the lattice constant $c_0$, i.e. one vortex between every double CuO$_2$ plane. Finally,
the top of the columns works as a grating with an area $A\approx \frac{H}{b}\Delta S$.

Diffraction of light from such a Josephson vortex lattice will appear effectively as a diffraction from a grating
made from a non-transparent material.
Our electrodynamic problem is to calculate the vector-potential created by the Josephson current,
Eq.~(\ref{jzdeltadelta}). To extract the contribution in the sum responsible for the
creation of the $n$-th diffraction maximum in the
reflected light we expand Eq.~(\ref{jzdeltadelta}) in Fourier series
\begin{equation}\label{jzFourier}
j_z(x)=\tilde{I}\sum_{n}\exp{\left\{i\left(k_x^{\mathrm{(in)}}+\frac{2\pi}{a}n\right)x\right\}}\,\delta(z)
A_{\omega,k;z}^{\mathrm{(in)}}e^{-i\omega t},
\nonumber
\end{equation}
where $ \tilde{I}=(\Delta S/4 \pi \lambda_c^2)(H/b) (c/a)$, because we substituted $\Delta S (H/b)$ for $\Delta S$ in
Eq.~(\ref{jzdeltadelta}) (see the reasoning in the previous paragraph).

The vector-potential created by the $n$-th current term in the
Fourier sum Eq.~(\ref{jzFourier}), which corresponds to the $n$-th diffraction
maximum, is
\begin{equation}\label{Aout}
A_z^{\mathrm{(out)}}=
\frac{2\pi \tilde{I}}{cK}e^{iqx+iKz-i\omega t}A_{\omega,k;z}^{\mathrm{(in)}},
\end{equation}
where $K=\sqrt{\frac{\omega^2}{c^2}-q^2}$ and $q=k_x^{\mathrm{(in)}}+\frac{2\pi}{a}n$.

From Eq.~(\ref{Aout}), and considering the light penetration depth to be of the
order of the light wavelength, $H\sim \lambda_{\mathrm{light}}$,
one finds the reflection coefficient for the $n$-th diffraction maximum,
consisting of three factors
\begin{eqnarray}\label{R}
R_{n}&=&\left|\frac{A^{\mathrm{(out)}}}{A^{\mathrm{(in)}}}\right|^2\\
&=&\left(\frac{c_0^2 \Gamma}{\lambda_c^2}\Gamma^2\right)^2
\left(\frac{\lambda_{\mathrm{light}}}{a}\right)^4f\left(n,
\theta^{\mathrm{(in)}},
\theta^{\mathrm{(out)}}\right).\nonumber
\end{eqnarray}
The first factor in Eq.~(\ref{R}) is intrinsic for the material. For
$\mathrm{YBa}_2 \mathrm{Cu}_3\mathrm{O}_7$ with a spacing between CuO$_2$ planes
$c_0\approx10$ \AA, an anisotropy parameter $\Gamma\approx 30$
and a London penetration depth in $z$-direction $\lambda_c\approx 1800$ \AA, it is
$$\left(\frac{c_0^2 \Gamma}{\lambda_c^2}\Gamma^2\right)^2\sim 1.$$
The second factor in Eq.~(\ref{R}) is the ratio of the light wavelength
and the Josephson vortex lattice constant, and
is adjusted by the external parameters the frequency of the falling light
and the applied magnetic field.  It should be of the order of unity,
diffraction is favorable when the wavelength of the scattered light is
compatible with the spacing between the columns of vortices
$\lambda_{\mathrm{light}}\sim a \,\,>\pi c_0 \Gamma$.
The third factor
$f\left(n,
\theta^{\mathrm{(in)}},
\theta^{\mathrm{(out)}}\right)$ is a function of the incident and the reflected angle of the $n$-th
diffraction maximum, and the effect can be best observed when it is of the order of unity,
when the angle $\theta^{\mathrm{(in)}}$ of the falling light is grazing, and  $\theta^{\mathrm{(out)}}$
corresponds to almost backscattering.

Actually the microscopic calculation using the Green's functions methods (see for example
the work by Pokrovsky and Pokrovsky\cite{PP} and the exercise book by Levitov and Shytov\cite{LS})
gives a $(\Delta/\hbar\omega)^2$ correction and logarithmic factors which are practically constants
(here  $\omega$ is the frequency of the falling light and $\Delta$
is the superconducting gap).
This decreases the magnitude of the effect, but we still have enough sensitivity to see the broad diffraction maximum.

\section{Conclusions}
For a qualitative consideration of the effect we used the static approximation.
A derivation by means of the anomalous Green's  functions reduces the diffraction reflection amplitude by
$(\Delta/\hbar\omega)^2$, but nevertheless the effect survives, and the intensity is not difficult to
measure. This result should encourage
experimentalists to consider the perspectives of such a simple experiment for the observation of a
new phenomenon -- light scattering by Josephson vortices.

\acknowledgments One of the authors (TMM) is thankful to Mark M\'esard
for the hospitality in LPTMS where this work was completed.


\begin{thebibliography}{99}

\bibitem{LD} W.~Lawrence and S.~Doniach,
    {\it Theory of layer structure superconductors}, Proc. 12th Inter. Conf.
    on Low Temperature Physics, Academic Pres of Japan, Kyoto, 1971, pp. 361-362.

\bibitem{LandauIX} E.M.~Lifshitz and L.P.~Pitaevskii,
    {\it Statistical Physics, Part 2,
    Landau and Lifshitz Course of Theoretical Physics, Volume 9}
    (Pergamon Press, Oxford, 1986).

\bibitem{Kopnin} N.~Kopnin,
    {\it Vortices in Type-II superconductors. Structure and Dynamics}
    (Oxford University Press, Oxford, 2001).

\bibitem{RevModPhys} G.~Blatter, M.V.~Feigel'man, V.B.~Geshkenbein, A.I.~Larkin, and V.M.~Vinokur,
    Rev. Mod. Phys. {\textbf 66}, 1125 (1994).

\bibitem{PP} S.V.~Pokrovsky and V.L.~Pokrovsky,
    {\it Plasma resonance in layered normal metals and superconductors},
    Proc. SPIE, \textbf{2157}, 93-110 (1994).

\bibitem{LS} L.S.~Levitov and A.V.~Shytov, {\it Green's functions. Theory and practice}
    (Fizmatlit, Moscow, 2003) (in Russian).

\end{thebibliography}
\end{document}